# Performance and Impairment Modelling for Hardware Components in Millimetre-wave Transceivers

Mythri Hunukumbure[1], Raffaele D'Errico[2], Antonio Clemente[2], Philippe Ratajczak[3], Ulf Gustavsson[4],Yinan Qi[1], Xiaoming Chen[5]

[1]Samsung Electronics R&D Institute UK, Staines, Middlesex TW18 4QE, UK
[2]CEA-Leti, Minatec Campus, 38054 Grenoble, France
[3] Orange Labs Sophia Antipolis, 905 rue A. Einstein, 06560 Valbonne France
[4] Ericsson Research, Lindholmspiren 11, 417 56 Gothenburg, Sweden
[5]Qamcom Research & Technology AB, Gothenburg 41285, Sweden

*Abstract*—This invited paper details some of the hardware modelling and impairment analysis carried out in the EU mmMAGIC project. The modelling work includes handset and Access Point antenna arrays, where specific millimeter-wave challenges are addressed. In power amplifier related analysis, statistical and behavioural modelling approaches are discussed. Phase Noise, regarded as a main impairment in millimeter-wave, is captured under two models and some analysis into to the impact of phase noise is also provided.

*Keywords—millimetre-wave, hardware impairments, antenna arrays, power amplifier, phase noise*

## I. INTRODUCTION

With the standardization process for 5G systems firmly under way now, there is more focus on the adaptation of millimeter-wave spectrum (loosely defined as 6-100 GHz) as a component of the overall 5G systems. While different radio systems have been employed in this spectrum for a number of years, adapting the spectrum for the specific needs of mobile communication brings multiple challenges. Some of these critical challenges have been addressed in the EU funded mmMAGIC project [1]. This paper presents some of the latest results in modelling performance and impairments in hardware components of the millimeter-wave transceiver chains.

Due to the need for high beam-forming gain to combat the excessive path loss in millimetre-wave frequencies, the design of antenna arrays takes special significance. We present results in modelling antenna arrays for handsets and access points (for both radio access and backhaul) in this paper. The power amplifier design in these frequencies is challenging as well, particularly in maintaining linear performance over wider bandwidths. Thirdly, we look at another hardware issue exacerbated in millimetre-wave frequencies, the phase noise. Two approaches for phase noise modelling and some analysis related to OFDM system design are discussed here.

The remainder of this paper is organized as follows. In section II, the millimetre-wave antenna array design is detailed, with results for handset and access point (AP) designs. The transmitarray design is introduced in this section as a potential AP design solution. Section III includes some insights into power amplifier nonlinearity and modelling approaches for millimetre-wave systems. The issue of phase noise is addressed in section IV, with the introduction of 2 phase noise models. The paper is concluded in section V, also with a look at potential future work in this area.

## II. MILLIMETRE-WAVE ANTENNA ARRAY MODELLING

### A. Antenna Designs for the Handset

Mm-wave antenna design for the handsets carry the usual problems of size and cost constraints and issues of antenna element coupling and the blocking effects with the hands. However the smaller inter-element distances make even the planar arrays possible within the handsets, to some degree.

To cover the full bandwidth (24.25-27.5 GHz), the classical printed dipole on ground plane has been chosen as elementary cell of the antenna array. This element is able to work in dual polarization in crossed dipole configuration. The radiation pattern for horizontal polarization at 26.0 GHz and the input impedance are presented below. For the 2 polarizations, the input impedances stay below -10 dB with around 9.0 dB gain.

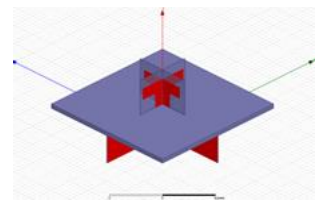

Fig.1: Elementary cell: crossed dipole on ground plane

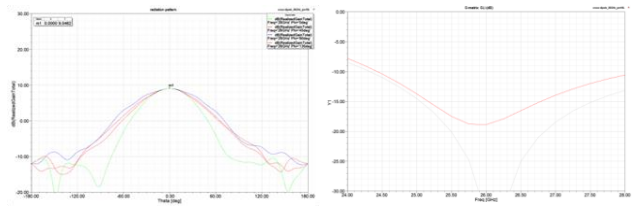

Fig.2: Radiation pattern and input impedances of the elementary cell

Considering the size of the element close to 0.5λ and the capability of the antenna array to scan the beam up to 60°, the spacing between elements of the array is equivalent to the size of the element. So we can observe a high level of coupling if the array is working in Horizontal and Vertical polarization that destroys the input impedance of each unit cell (0), a ±45° polarization allow to keep 0.5λ spacing thanks to a lower level of coupling, as shown in Fig. 3.

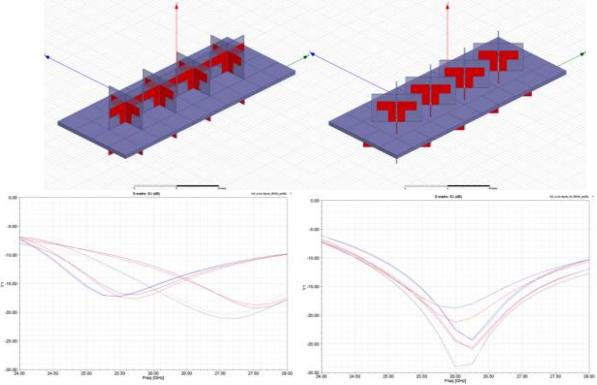

Fig.3: input impedances for 2 polarization configurations: H&V (right) and ±45° (left)

Several radiation patterns for main beam are presented for 1x4, and 1x8 elements respectively in Fig. 4 and 5. The gains are summarized for the different configurations and scanned angles in Table 1.

From all the configuration and scanned angles (0 to 60° with a 10° step), we can observe that the higher gain is not obtained for the 0° angle configuration but for a scanned angle around 20° taking into account the polarization of the element, the size of the ground plane, and the coupling between each elements of the array.

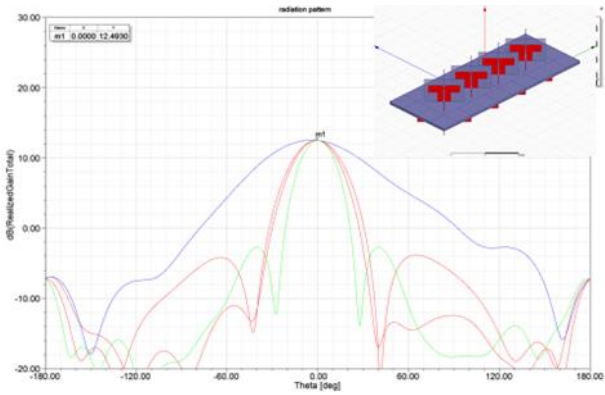

Fig. 4: 1x4 elements array – radiation pattern at 26.0 GHz.

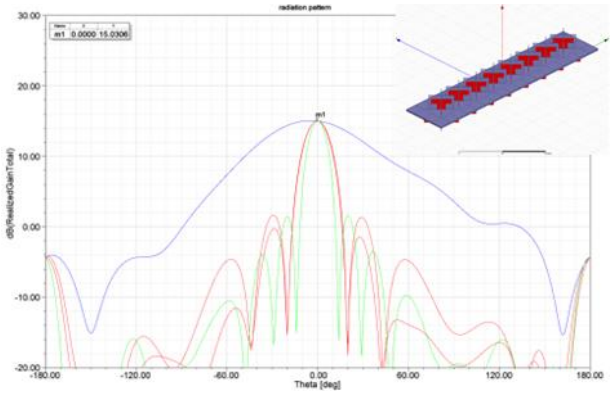

Fig. 5: 1x8 elements array – radiation pattern at 26.0 GHz.

TABLE 1- ARRAY ANTENNA GAINS

| array configurations | Gain (dBi) | |
|---|---|---|
| | *scanned angle < 30°* | *scanned angle = 60°* |
| 1x4 | 13.0 | 10.0 |
| 2x4 | 14.3 | 11.0 |
| 4x4 | 16.9 | 12.8 |
| 1x8 | 15.7 | 10.6 |

Equivalent gain can be obtained with a lower number of elements (6x6 instead of 8x8) with a larger spacing (0.7 l). The drawback is a reduced scanned angle to 25° to avoid grating lobes but a rearrangement of the array by destroying the array periodicity allows to enlarge this limit as presented on Fig. 6.

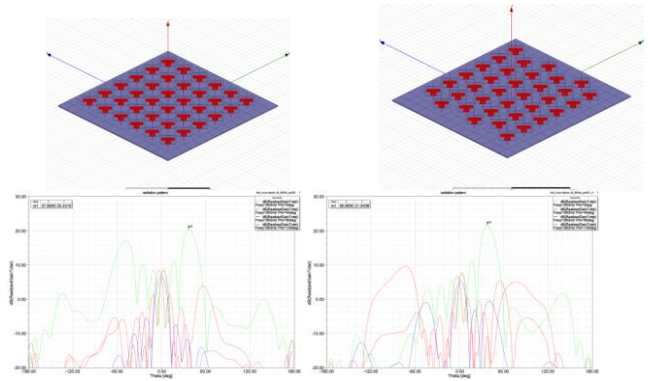

Fig. 6: 6x6 elements arrays – radiation pattern at 26.0 GHz: periodic (left) and non periodic in the horizontal plane (right).

*B. Antenna designs for Access and Backhaul at AP*

5G access point includes two functionalities: the radio access link, which guarantee the bidirectional link between users and access point and the backhaul/fronthaul link connecting one access point to the core-network or to a common base band unit. Typical antenna specifications for both applications are presented in Table 1. In general, for radio access high gain (> 20 dBi) antennas with analogue or hybrid beamforming capability are needed to manage both multi-users and mobility. In the case of backhaul/fronthaul link, a gain > 30 dBi and fixed beam or a limited scanning capability (±10° on one plane) are required. In fact, for this kind of link, the beam-steering could be used to implement self-alignment function. Others practical constraints for the access point antennas are the limited antenna size, cost and complexity.

TABLE 2- TYPICAL ACCESS POINT ANTENNA SPECS AT K-BAND

| Specifications | Radio access | Backhaul/fronthaul |
|---|---|---|
| Typ. bandwidth (GHz) | 3-4 GHz | 3-4 GHz |
| Typ. gain (dBi) | 20 - 25 | 30 – 35 |
| Polarization | Linear/Dual linear | Linear/Dual linear |
| Radiation mask | - | ETSI [3] |
| Beam-forming | Analogue/hybrid | Fixed/switched beam |
| Beam-steering | ±60° (2D spatial window) | ±10° (on one plane) |

Transmitarrays (Fig. 7) based on standard printed circuit board (PCB) technology are excellent candidates for millimetre-wave access point antenna implementation (for both radio access and backhaul/fronthaul). In fact, at millimetre-wave frequencies, the use of a spatial feeding (illumination of the planar array using a focal source) is extremely attractive in terms of loss if compared to standard phased arrays, which suffer of relatively high insertion loss in the feeding network. In addition, electronically reconfigurable transmitarray with beam-forming can be easily implemented by integrating active devices on the unit-cell of the flat array [4].

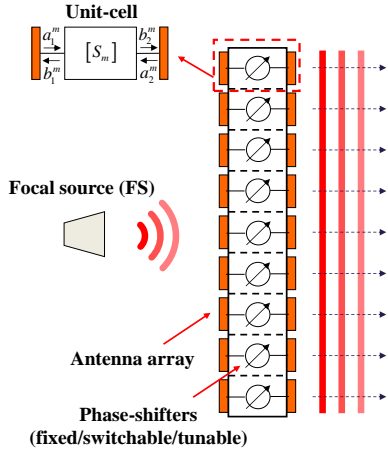

Fig. 7: Schematic view of a transmitarray antenna.

Realistic transmitarray antennas parameters and performance for radio access and backhaul/fronthaul in the K-band (24.25 – 27.5 GHz) are presented in Table 3 and 4, respectively. The proposed transmitarray are illuminated by a standard gain horn with a gain of 10 dBi and the unit-cells characteristics have been extrapolated by considering designs in Ka-band [4]. The results have been extracted by using an *in-house* simulation tool based on analytical formulas and realistic data of the focal source and unit-cells, from electromagnetic simulation. In the case of the transmitarray for the radio access, two designs, considering 1- and 2-bit phase quantization, based on the electronically reconfigurable unit-cell [4], have been considered.

The beam-steering capability of the 1-bit 20×20 unit-cell transmitarray and of the 2-bit 14×14 unit-cell transmitarrays are presented in Fig. 7. Two and four p-i-n diodes are used on each unit-cell to locally control the phase-shift on the transmitarray aperture in the case of 1- and 2-bit phase quantization, respectively. Phase quantization is implemented to make a trade-off between cost, complexity and performances (bandwidth and insertion loss) of the electronically reconfigurable unit-cell. Hybrid beam-forming architecture could be easily implemented by adding additional focal sources.

TABLE 3: TYPICAL TRANSMITARRAY ANTENNA PARAMETERS FOR RADIO ACCESS AT K-BAND

| Parameter | 1-bit | 2-bit |
|---|---|---|
| Number of UC | 20×20 | 14×14 |
| UC size (mm²) | 5×5 | 5×5 |
| Array size (mm²) | 100×100 | 70×70 |
| Number of phase states | 2 | 4 |
| Relative phase-shift | 180° | 90° |
| Number of p-i-n diodes | 800 | 784 |
| Focal distance (mm) | 60 | 45 |
| FS gain (dBi) | 10 | 10 |
| Gain (dBi) | 23.7 | 23.4 |
| Total loss (dB) | 2.5 | 2.9 |

In the case of backhaul/fronthaul applications at K-band, the analyzed transmitarrays are based on a passive unit-cell with 3-bit phase quantization. The unit-cell realistic characteristics have been extrapolated from our previous works. The antenna parameters and radiation patterns compared to the ETSI specification for point-to-point communication links computed as a function of the phase quantization (1-, 2-, and 3-bit) are presented in Table 3 and Fig. 9. Also in this case beam-switching can be easily performed by using a multi focal source system.

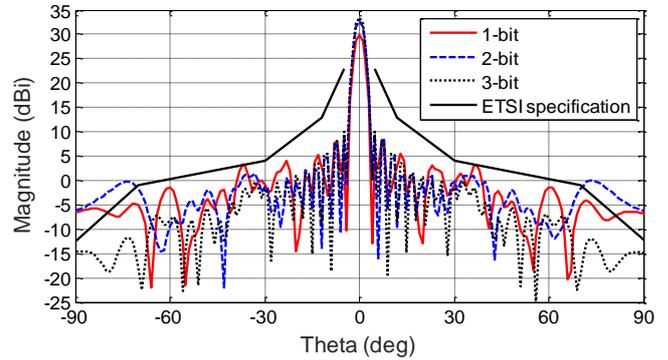

Fig. 8: Simulated radiation patterns of fixed beam transmitarray for backhaul/fronthaul computed at the central frequency as a function of the phase quantization

TABLE 4: TYPICAL TRANSMITARRAY ANTENNA PARAMETERS FOR BACKHAUL/FRONTHAUL AT K-BAND

| Parameter | 1-bit | 2-bit | 3-bit |
|---|---|---|---|
| Number of UC | 40×40 | 40×44 | 40×44 |
| UC size (mm²) | 5×5 | 5×5 | 5×5 |
| Array size (mm²) | 200×200 | 200×200 | 200×200 |
| Number of phase states | 2 | 4 | 8 |
| Relative phase-shift | 180° | 90° | 45° |
| Focal distance (mm) | 134 | 134 | 134 |
| FS gain (dBi) | 10 | 10 | 10 |
| Gain (dBi) | 30.5 | 33.0 | 33.5 |
| Total loss (dB) | 1.5 | 1.5 | 1.5 |

## III. POWER AMPLIFIER MODELLING

### A. Behavioral modeling for large arrays

In general, the area of power amplifier modeling is rather mature relying on decades of research in non-linear systems models. The Volterra-series framework, [5], have provided a solid ground for power amplifier modeling from which specializations such as the generalized memory polynomial (GMP), [6], has arisen.

One major challenge with moving toward mm-wave frequencies for wireless communication is to generate transmit power. As dictated by the Johnson limit [7], the output power capability decreases proportionally to the squared operating frequency. Thus, large antenna arrays are not only beneficial, but rather necessary in order to provide sufficient link-budget. When introducing power amplifiers in environments such as large, dense antenna array's, they become subject of mutual coupling which introduces a new source of distortion not covered in the regular Volterra-based framework. For this reason, development in the field of power amplifier modeling for antenna arrays has flourished the last couple of years with some progress based on a GMP basis, in which the mutual coupling effects is modeled via a secondary variable. This modeling framework has been widely used in power amplifier distortion evaluations and its impact on massive MIMO [8].

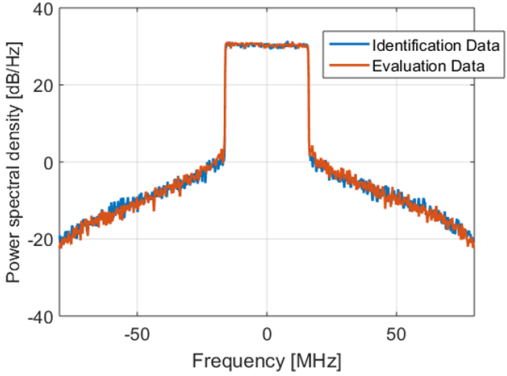

Figure 9: - Measured and simulated power spectral density using a GMP model for a 20 MHz LTE carrier. The model is using a 7$^{th}$ order non-linearity, 5 memory-taps and 2 cross-terms.

### B. Statistical modeling for link- and network analysis

For certain applications such as link- or network-level evaluations, it may be practical to limit the modeling to knowledge about the statistical properties to the power amplifier distortion.

One of the simplest approaches is to model the impairments as a static, multiplicative gain and phase error, which allows for simple approximate SINR evaluation through analytical manipulation. A more involved approach based upon computing the distortion covariance matrix. The entries in the distorted transmit covariance matrix can be developed from the corresponding behavioral model. We will exemplify this using a single-antenna system here. Using a simple, static third order polynomial, written as

$$y = \theta_1 x + \theta_2 x|x|^2 \quad (1)$$

we can use the Bussgang theorem to decompose the model into a first order stochastic approximation which is written as

$$y = \alpha x + w \quad (2)$$

The parameters are defined and computed for the third order polynomial case as

$$\alpha = \frac{E[y^*x]}{\sigma_x^2} = \theta_1 + 2\theta_2 \sigma_x^2 \quad (3)$$

$$\sigma_w^2 = E[\|y - \alpha x\|^2] = 2|\theta_2|^2 (3\sigma_x^6 + 2\sigma_x^8) \quad (4)$$

It is interesting to notice that the distortion term 1) does not depend on the linear term in the model, and 2) grows cubically with transmit power, $\sigma_x^2$.

In the multi-antenna case, these parameters need to be computed across the array and the corresponding model can be formulated as

$$\mathbf{Y} = \mathbf{\Lambda X} + \mathbf{W} \quad (5)$$

where $\mathbf{X} \sim CN(\mathbf{0}, \mathbf{C_{xx}})$, $\mathbf{W} \sim CN(\mathbf{0}, \mathbf{C_{ww}})$ and $\mathbf{\Lambda} = \text{diag}(\alpha_1, \dots, \alpha_M)$.

## IV. MODELLING OF PHASE NOISE

The spectrum of an ideal oscillator should be a Dirac delta function. In reality, due to various noises in the oscillator, the spectrum of an oscillator spread out around the carrier frequency, resulting in common phase error (CPE) and inter-carrier interference (ICI). The ICI effect can be alleviated by having larger subcarrier spacing, as in wireless local area network (WLAN) system. The CPE causes a phase rotation for all the subcarrier symbol in an orthogonal frequency division multiplexing (OFDM) symbol, which can significantly degrade the performance of the OFDM system (if left uncompensated).

### A. Detailed PN model and Analysis

The phase noise of a phase-locked loop (PLL) based oscillator consists of three main noise sources, i.e., noises from the reference oscillator $\theta_{\text{ref}}$, the phase-frequency detector and the loop filter $\theta_{\text{LP}}$, and the voltage controlled oscillator (VCO) $\theta_{\text{VCO}}$, as shown in Fig. 1. The Laplace transform of the phase noise of the PLL-based oscillator is given as [10]10

$$\theta_{\text{out}}(s) = \frac{N_D K_{\text{VCO}} Z(s)(\theta_{\text{ref}}(s) + K_D \theta_{\text{LP}}(s)) + sN_D \theta_{VCO}(s)}{sN_D + K_D K_{\text{VCO}} Z(s)} \quad (6)$$

where $K_D$ denotes the gain of the phase-frequency detector, $K_{\text{VCO}}$ represents the sensitivity of the VCO, $Z(s)$ represents the loop filter, and $1/N_D$ is the frequency divider. The detailed modeling parameters are listed in Table 4-2 of [10]10. As an example, Fig. 1 shows the estimated power spectral densities (PSDs) of the phase noises (in the "high" mode [10]10) at 5, 28, and 60 GHz, respectively.

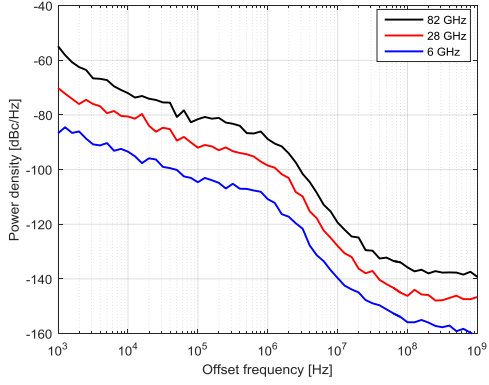

Fig. 10: PSD of the phase noise.

The frequency domain signal of a multiple-input multiple-output (MIMO) OFDM system is given as

$$\mathbf{y} = (\mathbf{G}_R \otimes \mathbf{I}_{N_R})\mathbf{H}(\mathbf{G}_T \otimes \mathbf{I}_{N_T})\mathbf{x} + \mathbf{w}. \quad (7)$$

Where $\mathbf{H}$ is an $N_R N \times N_T N$ block diagonal channel matrix whose $k$th diagonal block entry $\mathbf{H}_k$ is the $N_R \times N_T$ MIMO channel matrix of the channel transfer functions at the $k$th subcarrier, $\mathbf{x} = \begin{bmatrix} \mathbf{x}_1^T & \mathbf{x}_2^T & \cdots & \mathbf{x}_N^T \end{bmatrix}^T$ is the $N_T N \times 1$ frequency-domain signal vector with $\mathbf{x}_k$ denoting the $N_T \times 1$ transmitted signal vector at the $k$th subcarrier, $\mathbf{y} = \begin{bmatrix} \mathbf{y}_1^T & \mathbf{y}_2^T & \cdots & \mathbf{y}_N^T \end{bmatrix}^T$ is the $N_R N \times 1$ frequency-domain signal vector with $\mathbf{y}_k$ denoting the $N_R \times 1$ received signal vector at the $k$th subcarriers, and $\mathbf{w}$ is a $N_R N \times 1$ additive white Gaussian noise (AWGN) vector $\mathbf{G}_T$ is phase noise contributions in the form of

$$\mathbf{G}_T = \begin{bmatrix} g_0^{Tx} & g_{N-1}^{Tx} & \cdots & g_1^{Tx} \\ g_1^{Tx} & g_0^{Tx} & \cdots & g_2^{Tx} \\ \vdots & \vdots & \ddots & \vdots \\ g_{N-1}^{Tx} & g_{N-2}^{Tx} & \cdots & g_0^{Tx} \end{bmatrix}, \quad (8)$$

whose elements are the discrete Fourier transform of the time-domain phase noise; and, analogously, the $(k, l)$th entry of $\mathbf{G}_R$ is $g_{(k-l)_N}^{Rx}$, with $(k-l)_N$ denotes $(k-l)$ mod $N$ [11].

Let $\mathbf{G}_T = g_0^{Tx}\mathbf{I}_N + \mathbf{P}_T$ and $\mathbf{G}_R = g_0^{Rx}\mathbf{I}_N + \mathbf{P}_R$, (3) can be rewritten as

$$\mathbf{y} = g_0^{Rx} g_0^{Tx} \mathbf{H}\mathbf{x} + \mathbf{e} + \mathbf{w}, \quad (9)$$

where $g_0^{Rx} g_0^{Tx}$ represents the CPE and $\mathbf{e}$ is the ICI term given as

$$\mathbf{e} = (\mathbf{P}_R \otimes \mathbf{I}_{N_R})\mathbf{H}(\mathbf{P}_T \otimes \mathbf{I}_{N_T})\mathbf{x} + g_0^{Rx}\mathbf{H}(\mathbf{P}_T \otimes \mathbf{I}_{N_T})\mathbf{x}$$
$$+ g_0^{Tx}(\mathbf{P}_R \otimes \mathbf{I}_{N_R})\mathbf{H}\mathbf{x}. \quad (10)$$

### B. A simple and effective PN model

A simpler yet effective phase noise model was also studied in the project, which brings certain advantages to the analysis of the impact of phase noise. The presented multi-pole/zero model is an extension to the single pole/zero model adapted for IEEE P802.15 [12]. With a few, carefully chosen poles and zeros, it was found that the phase noise spectra of practical oscillators can be effectively matched. The power spectral density behaviour of the proposed model is given by the following equation:

$$S(f) = PSD_0 \prod_{n=1}^{N} \left( \frac{1+\left(\frac{f}{f_{z,n}}\right)^2}{1+\left(\frac{f}{f_{p,n}}\right)^2} \right) \quad (11)$$

The model gives following practical advantages:

- Practical phase noise power spectra can be well approximated with a few pole/zeros, as it gives more flexibility than a single pole/zero model. The challenge is to identify the correct poles and zeros to suit to a practical oscillator/ frequency synthesizer.

- Provides an easy framework to convert the PSD of analog phase noise to that of discrete-time phase noise (i.e., baseband version) for simulation by using the bilinear transform with given pole/zeros. This is transforming the s-domain multi-pole/zero function to the z-domain.

Table 5 shows two parameter sets which are obtained from practical oscillators operating at 30GHz and 60GHz, respectively. We call them "Set-A" and "Set-B" for simplicity.

Table 5: Example of parameter sets for the proposed PN model

|  | Parameter Set-A | Parameter Set-B |
|---|---|---|
| Carrier frequency ($f_{c,base}$) | 30 GHz | 60 GHz |
| PSD0 (dBc/Hz) | -79.4 | -70 |
| Fp (MHz) | [0.1, 0.2, 8] | [0.005, 0.4, 0.6] |
| Fz (MHz) | [1.8, 2.2, 40] | [0.02, 6, 10] |

Fig.11 shows the power spectral densities for 3 carrier frequencies with these parameters. If the operating carrier frequency is changed, the PSD is shifted by $20\log_{10}(f_c / f_{c,base})$ dBc/Hz.

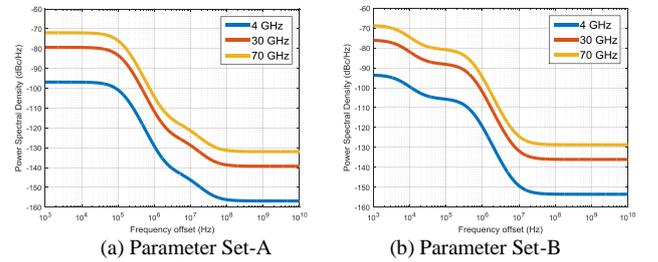

(a) Parameter Set-A  (b) Parameter Set-B

Figure 11: Phase noise power spectral density

#### 4.1.1 Phase Tracking Reference Signal (PTRS) design

PTRS design is an active topic in 3GPP NR standardisation work [13]. The PTRS are known pilot symbols, inserted into the radio sub-frame at the transmitter, so the receiver can correct the Common Phase Error (CPE). The CPE is a main component of PN and it rotates the symbol constellations across all sub-carriers in an OFDM system by an equal amount. There is an obvious trade-off in the allowable density of PTRS, w.r.t the acceptable BLER (block error rate) and the pilot overheads.

Using the above PN model, the CPE was synthesized in an OFDM based 5G NR transmission and the performance for different PTRS densities was evaluated. In 3GPP NR terminology the Physical Resource Block (PRB) occupies 12 sub-carriers in the frequency domain and 7 symbols in the time domain. PRB is the basic unit of resource allocation to the user and multiple PRBs can be allocated, depending on the data rate of the supported user. Assuming that the user is allocated 100 PRBs (equivalent to 20MHz BW) the performance of different PTRS densities is illustrated in Figure 12. The performance of similar frequency domain PTRS densities but when the user is having different PRB allocations is shown in Figure 13.

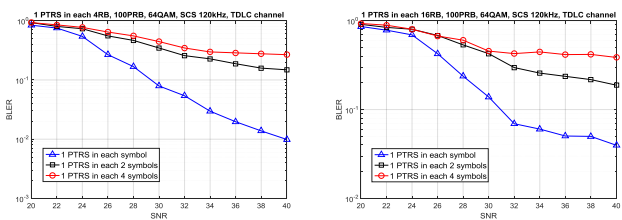

Figure 12: BLER results for PTRS densities for a 100 PRB allocation

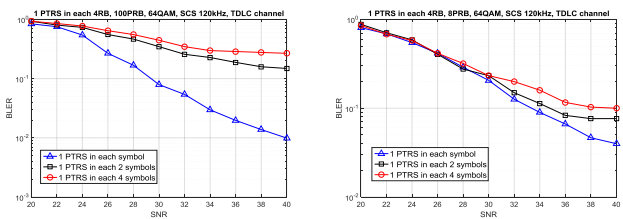

Figure 13: BLER results for PTRS densities for different PRB allocations

The results in Fig. 12 show that the BLER performance degrades when both the frequency domain (PTRS in sub-carrier per 4 or 16 PRBs) and time domain (PTRS in each or each 2 or each 4 symbols) densities are reduced. There is a bigger performance gap from 1 to 2 symbol densities than 2 to 4 symbol densities in both plots. In Fig. 13, the performance is less sensitive to time domain PTRS density when the user is allocated a narrower bandwidth (less PRBs). The overall results indicate that the time domain PTRS density needs to be virtually every symbol, but in the frequency domain, PTRS can be in a sub-carrier per every few PRBs. This work was also submitted to 3GPP RAN1 [13] and discussed in the meeting.

V. CONCLUSIONS

The mmMAGIC project has made significant strides in understanding the millimeter-wave transceiver performance and non-ideal behavior through analytical and modelling based studies. This paper presents some of the current results from antenna array, power amplifier and phase noise modelling and impairment impact analysis.

The general conclusion from this body of work is that while the transceiver hardware challenges are significant, they can be overcome with proper analysis and design solutions. Some of the antenna solutions are discussed in this paper. The project will continue to research these topics and will provide the final results by the end of June 2017.

ACKNOWLEDGMENT

The European Commission funding under H2020- ICT-14-2014 (Advanced 5G Network Infrastructure for the Future Internet, 5G PPP), and project partners: Samsung, Ericsson, Aalto University, Alcatel-Lucent, CEA LETI, Fraunhofer HHI, Huawei, Intel, IMDEA Networks, Nokia, Orange, Telefonica, Bristol University, Qamcom, Chalmers University of Technology, Keysight Technologies, Rohde & Schwarz, TU Dresden are acknowledged.